\documentclass[a4paper,11pt]{article}
\pdfoutput=1 

\usepackage{jinstpub} 

\usepackage{lineno}

\usepackage{multirow}

\title{\boldmath Study of polysterene based scintillator ageing in the DANSS experiment}

\author[a,b,1]{I.~Alekseev\note{Corresponding author.},}
\author[c]{V.~Belov,}
\author[b,c,d]{A.~Bystryakov,}
\author[b]{M.~Danilov,}
\author[c]{D.~Filosofov,}
\author[c]{M.~Fomina,}
\author[e]{P.~Gorovtsov,}
\author[a,f]{Ye.~Iusko,}
\author[c]{S.~Kazartsev,}
\author[a,b,e]{A.~Kobyakin,}
\author[a,b,e]{A.~Krapiva,}
\author[c]{A.~Kuznetsov,}
\author[a,b,e]{N.~Mashin,}
\author[g]{I.~Machikhiliyan,}
\author[c]{D.~Medvedev,}
\author[a,b]{V.~Nesterov,}
\author[b,c]{D.~Ponomarev,}
\author[c]{I.~Rozova,}
\author[a,c,d]{N.~Rumyantseva,}
\author[a,b]{V.~Rusinov,}
\author[c]{A.~Salamatin,}
\author[a,b]{E.~Samigullin,}
\author[c]{Ye.~Shevchik,}
\author[c]{M.~Shirchenko,}
\author[h]{Yu.~Shitov,}
\author[a,b]{N.~Skrobova,}
\author[a,b]{D.~Svirida,}
\author[a]{E.~Tarkovsky,}
\author[b,e]{A.~Yakovleva,}
\author[c]{E.~Yakushev,}
\author[c]{I.~Zhitnikov}
\author[c,i]{and D.~Zinatulina}

\affiliation[a]{NRC "Kurchatov Institute",\\ Academician Kurchatov Square 1, Moscow, 123182, Russia}
\affiliation[b]{Lebedev Physical Institute of the Russian Academy of Sciences, \\ Leninskiy avenue 53, Moscow, 119991, Russia}
\affiliation[c]{Joint Institute for Nuclear Research, \\ Joliot-Curie str. 6, Dubna, Moscow region, 141980, Russia}
\affiliation[d]{Dubna State University, \\ Universitetskaya St. 19, Dubna, Moscow Region, 141980 Russia}
\affiliation[e]{Moscow Institute of Physics and Technology, \\ Institutskiy lane 9, Dolgoprudny, Moscow Region, 141701, Russia}
\affiliation[f]{National Research Nuclear University MEPhI, \\ Kashirskoe shosse 31, Moscow, 115409, Russia}
\affiliation[g]{Federal State Unitary Enterprise Dukhov Automatics Research Institute, \\ Sushchevskaya str. 22, Moscow 127055, Russia}
\affiliation[h]{Institute of Experimental and Applied Physics, Czech Technical University in Prague\\ Husova 240/5, Prague, 110 00 Czech Republic}
\affiliation[i]{Voronezh State University, \\ Universitetskaya square 1, Voronezh, 394018, Russia}
\emailAdd{igor.alekseev@itep.ru}

\abstract{DANSS is a spectrometer of reactor antineutrino based on plastic 
scintillator. The sensitive volume of the detector is made of 2500 polystyrene based
scintillator plates with wavelength shifting (WLS) fiber readout (strips). We present a study 
of the strips light yield during 6.5 years of DANSS continuous running. Overall ageing
at the rate $0.55 \pm 0.05 (\mathrm{syst.})$~\% per year is observed  that is considerably smaller than in other similar experiments. We also observe the WLS fiber attenuation length
shortening at the rate $0.26 \pm 0.04(\mathrm{stat.})$~\% per year.}


\begin{document}
\maketitle
\flushbottom

\section{Introduction}
The idea of the DANSS detector~\cite{DANSS_JINST, DANSS_PLB} was to have a scintillation neutrino spectrometer on 
a movable platform as close as possible to the core of the industrial nuclear reactor. The site just below the
reactor body provides several advantages like very high neutrino flux ($\sim 5\times 10^{13}\bar{\nu}_e/$cm$^2/$s) and
moderate overburden ($\sim 50$~m w.e.), but at this site only low flammable materials are allowed due to the fire safety reasons. 
This justified to assemble the sensitive volume from polystyrene based solid state scintillator. 
Continuous operation of the experiment since October 2016 in a climate controlled room provides 
a very good data sample to study ageing effects of the detector. 

Ageing of the plastic scintillators is already known for many years~\cite{Barnaby1962}. There are several 
mechanisms which are usually considered. An excess mechanical stress at a level above elastic and
anelastic deformation results in nonrecoverable damage to the scintillator structure like shear plasticity and 
crazing~\cite{Kambour1973}, creating obstacles on the passage of the light. Humidity of the environment 
can penetrate into the bulk of the scintillator and produce fogging effect in it~\cite{Sword2017}. 
Atmosphere oxygen together with ultraviolet radiation could develop photolysis of the polystyrene 
resulting in its yellowing~\cite{Yousif2013}. All of these effects affect transparency of the
plastic and result in smaller amount of scintillation light collected. Plastic exposition
in the DANSS experiment was very mild avoiding mechanical stresses, excessive humidity and
ultraviolet radiation.

This work was inspired by the recent study performed by T2K experiment~\cite{T2K}. The T2K paper covers 10 years
of the operation and reports light yield decrease in the range of $0.9 \div 2.2$~\% per year 
for different subsystems of the detector. Measured degradation of polystyrene
scintillator strips from the MINOS experiment~\cite{MINOS}, which are similar to ours, is 2~\% per year.
Though ageing as large as $7 \div 10$~\% per year was observed in MINERvA~\cite{MINERvA} at $80^{\circ}$F 
($27^{\circ}$C).

\section{Experimental conditions}
The DANSS experiment \cite{DANSS_JINST} is located at the $4^{th}$ reactor unit of the Kalininskaya NPP about 350~km NW 
from Moscow. The detector is placed in a room just below the core of the reactor. For the sake of the electronics 
cooling, the room is air conditioned; thus, the temperature around 20 $^{\circ}$C is maintained all the time.
The relative humidity in the room is also very low, below the lower limit of 10\% of the humidity
detector used as the environment monitor. Despite of the close vicinity of the reactor, the room is clean in terms of
radiation background and corresponds to a laboratory room environment.

\begin{figure}[ht]
\centerline{\includegraphics[width=0.6\textwidth]{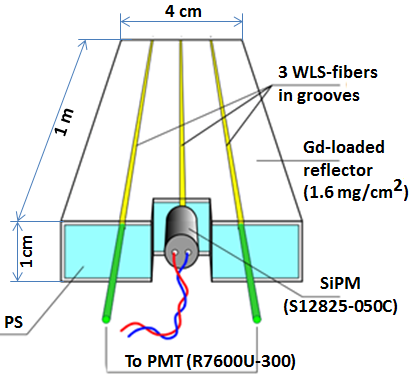}}
\caption{Scintillation strip layout.\label{fig:strip}}
\end{figure}

The detector sensitive volume is built from scintillator plates (strips) which were extruded from polystyrene by the Institute of 
Scintillating Materials, Kharkiv, Ukraine. The strip design was developed especially for this experiment with very specific
material composition and manufacture procedure~\cite{RepStrip}. The core plastic is doped with 1\% PPO 
(2,5-Diphenyloxazole) and 0.03\% POPOP (1,4-bis(5-phenyloxazol-2-yl) benzene) fluors. 
The surface is covered by about $\approx$0.3~mm co-extruded diffuse reflection layer. The layer consists of polystyrene with 18\%$_\mathrm{wt}$ admixture of
rutile TiO$_2$ and 6\%$_\mathrm{wt}$ of gadolinium oxide Gd$_2$O$_3$. Gadolinium has natural isotope composition and 
is used to capture neutrons from the inverse beta-decay after their moderation. The final gadolinium density is about 1.6~mg/cm$^2$, 
which corresponds to $\sim$0.35\%$_\mathrm{wt}$ with respect to the whole detector body. The layout of the strip is shown in Figure~\ref{fig:strip}. 
The strip has 1~m length, 40~mm width and 10~mm thickness. There are three grooves on the top surface of the strip with wavelength shifting (WLS)
fibers Y-11(200)M by Kuraray~\cite{KURARAY}. The fibers are glued into groves with a two component siloxane based optical gel from SUREL SUREL-SL1
(St. Petersburg, Russia)~\cite{SUREL}.
The central fiber is read out by a SiPM S12825-050C by Hamamatsu. Two side fibers from each of the 50 
neighbor strips are collected into 100-fiber bundles and led to a conventional PMT R7600U-300 by Hamamatsu. 
The opposite ends of the fibers are polished and covered by reflective paint ``Silver shine'' M415001K00250 by Manoukian Argon (Italy). 
The strips are put in rows of 25 each. There are
100 rows, which form a cubic meter of the sensitive volume of the detector. Strips in adjacent rows are perpendicular to each other. 
The strips are supported by copper plates, which serve as an inner layer of the passive shielding. The copper layer is 5~cm thick on 
average. The aim during the assembly was to have most of the weight residing on the copper frame, though there is some sag of the strips 
in their middle and some touching between the strips could not be excluded.
The outer part of the shielding is assembled from 8~cm of borated polyethylene, 5~cm of lead and another 8~cm of borated
polyethylene. The shielding significantly decreases both gamma and neutron backgrounds, making radiation doses in the strip location well below
common room conditions. Three centimeter thick scintillator veto counters cover the detector outside the shielding on all sides with the exception of the
bottom and serve for tagging muons. The front-end electronics are placed on two sides of the cube in a slit between copper and borated polyethylene.
The electronics are cooled by a chilled water cooling system, providing a stable temperature for the central part of the detector.
Unfortunately, the temperature at the SiPM location varies from 18 $^{\circ}$C for top and bottom rows to 23 $^{\circ}$C for middle rows.
DANSS data acquisition system (DAQ) is based on 64-channel, 12-bit, 125 MSPS waveform digitization modules (WFD) 
\cite{WFD}, which sample analog signals from photodetectors in 512~ns windows.

The SiPM bias voltages were set once at the very beginning of the detector operation and fixed for the rest of the running period. 
The setting was carried out based on a single pixel signal with the aim of about 12 ADC units per pixel. The voltages were
fixed once, and no voltage adjustment for temperature changes or any other reason was ever carried out. 
The DAQ operated with a common trigger, fired when either an instant digital sum of all PMTs was above a 0.5~MeV threshold or
a veto system was triggered. In addition to the common trigger, each SiPM channel was triggered individually once per
50,000 pulses in it. These individual triggers were used to collect the SiPM thermal noise data, out of which single pixel charge and
crosstalk were extracted. Such calibration was performed for each SiPM approximately once per half 
hour~\cite{CalibPaper}. For the neutrino analysis stream, the energy scale of all detectors 
is calibrated by atmospheric muons approximately every two days. 
The present analysis does not use the energy scale from the main analysis stream and
takes SiPM data only in the form of number of photo carriers (ph.c.) per hit. 
This number is corrected for SiPM saturation caused by the finite number of pixels.
SiPM S12825-050C has 667 pixels in its nearly square matrix. If more than one photon triggers a single pixel, a signal equiv{a}lent of
only one pixel is produced. Only $511.6 \pm 19.1$ of SiPM pixels are effectively illuminated by the WLS fiber 
in DANSS strips~\cite{CalibPaper}. This number was used for the saturation correction.
This paper is based on the statistics collected during the period from October 10, 2016 till March 15, 2023.
 
\section{Analysis}
\begin{figure}[ht]
\centerline{\includegraphics[width=\textwidth]{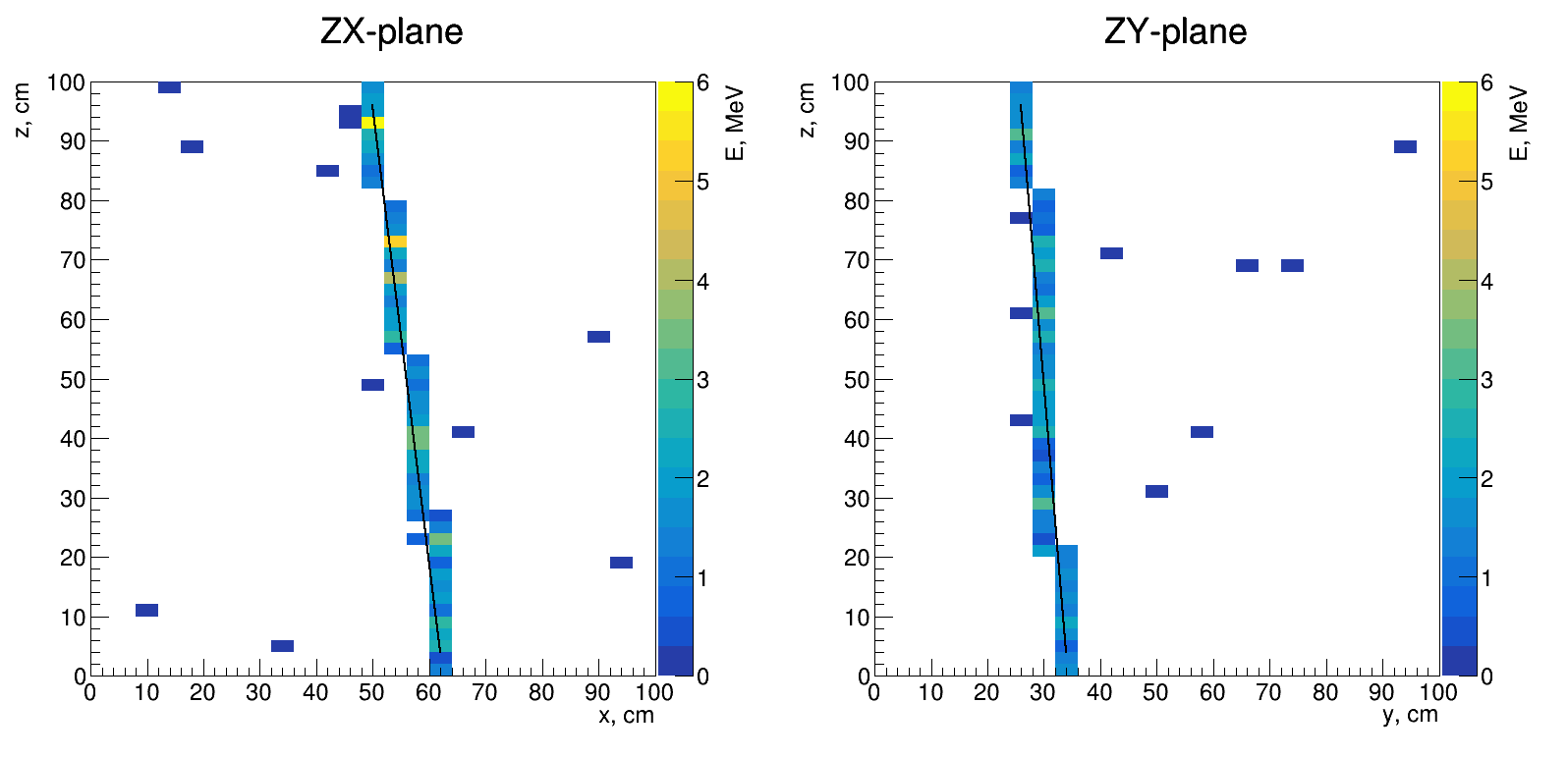}}
\caption{An example of selected muon event. The color scale corresponds to strip energy
deposit. A black line shows muon track position reconstructed by its ends.\label{fig:event}}
\end{figure}

Though the detector has notable overburden, about 40 atmospheric muons pass each second through the detector~\cite{DANSS_muon}.
Energy deposit from close to vertical muons is used as a standard candle. We start with selection
of triggers with vertical muons. The event should have number of hits in the above 90. 
This effectively selects events with muons passing through the most of the detector. Track
coordinates in the lowest and the highest 8 planes are calculated by averaging hits with a weight proportional to their energy deposit. 
A muon track is defined as a line connecting the points with these averaged coordinates. In order
to select vertical muons only events with distances below 20 cm in the horizontal plane between
these points are selected. So the tangent of the track zenith angle is below 0.2 and the difference 
between lengths of track in a strip is less than 2\%. An example of a selected event is shown in
Figure~\ref{fig:event}. A hit energy is shown by a color and the line corresponding to the track
is shown also. Dark blue rectangles show low energy hits, produced by SiPM noise.

At the next stage hits are selected. Several criteria are used:
\begin{itemize}
    \item distance from muon track is less than 1.5 strip width;
    \item there is no hit in the neighbor strips of the same layer. This is done to avoid
    partial track segments shared between adjacent strips;
    \item there are hits in the strips below and above in the same projection.
\end{itemize}
In order to have more narrow distributions the value of ph.c. is corrected for longitudinal 
profile along the strip. We use average profiles extracted from the data as described in~\cite{CalibPaper}.
Only strips which were alive throughout the whole period and all the time connected to the same DAQ channel were selected. 
This requirement made a significant cut over available channels leaving only 1713 out of 2500 for further analysis,
but it gives a valuable simplification of the analysis.

\begin{figure}[ht]
\centerline{\includegraphics[width=0.6\textwidth]{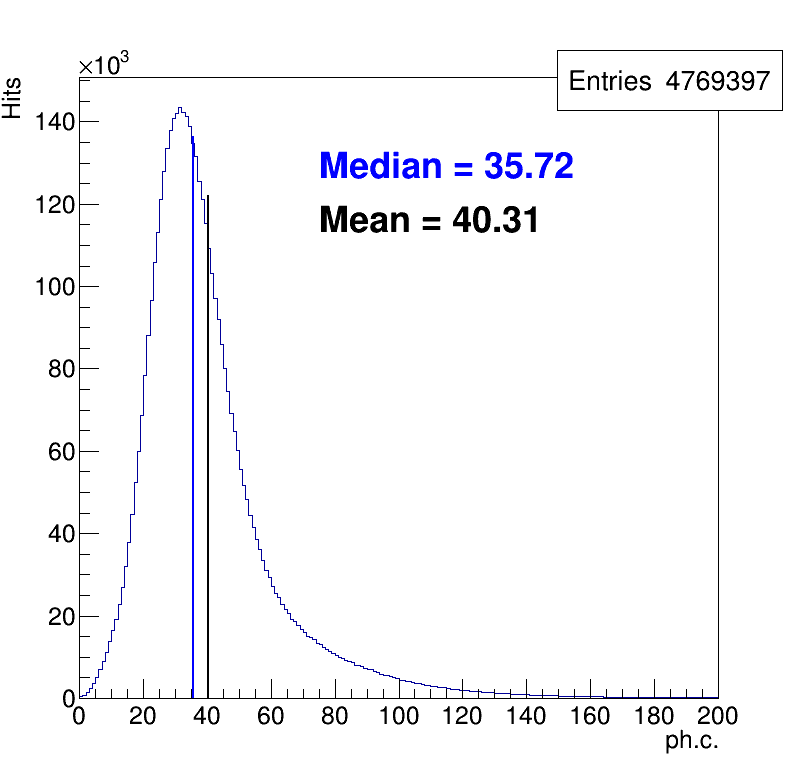}}
\caption{An example of number of photo carriers per hit distribution.\label{fig:median}}
\end{figure}

The whole data was splitted into 100-files chunks, which correspond to approximately 2 
days of data taking. A distribution of number of photo carriers per hit for all hits in 
some data chunk is shown in Figure~\ref{fig:median}. Median value is used in the analysis 
as a measure of the light yield (LY). 

\section{Results}

\begin{figure}[ht]
\includegraphics[width=\textwidth]{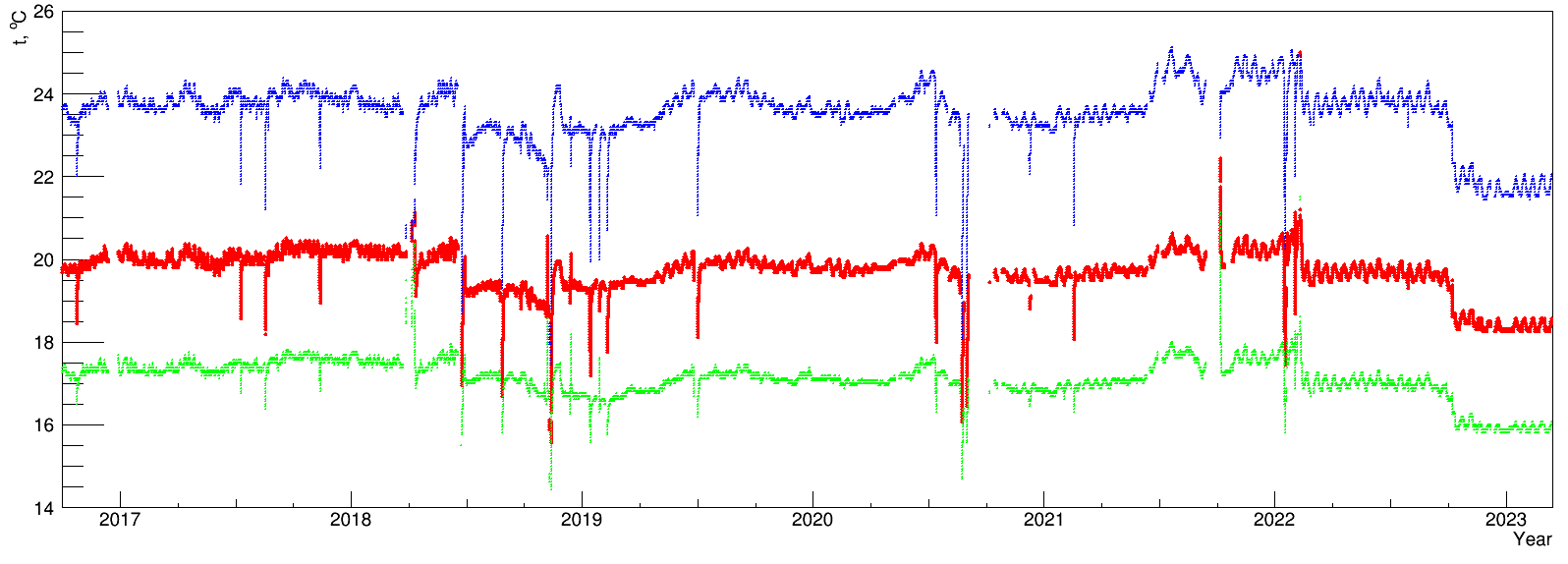}\\
\includegraphics[width=\textwidth]{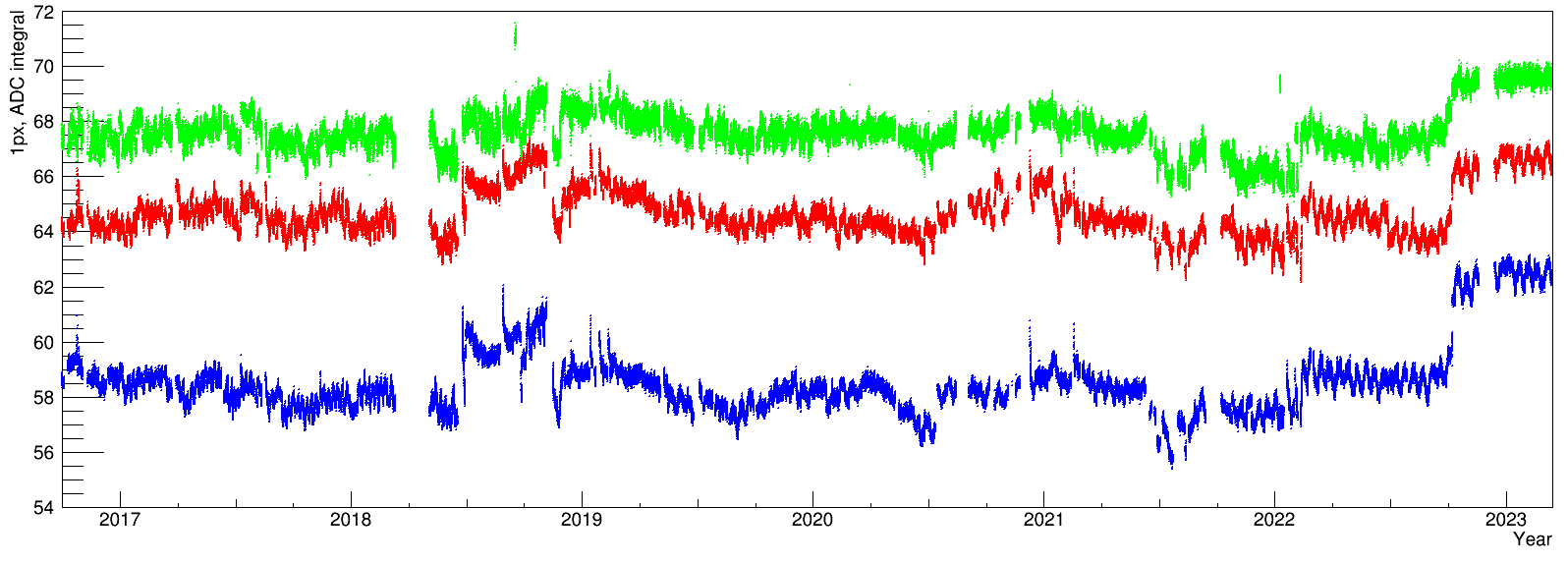}
\caption{Temperature dependence as a function of calendar time (the top panel). Different colors show 3
sensors placed in the center (blue), right top corner (red) and left bottom corner (green) of the "Y" readout plane. The 
calendar time dependence of the single pixel signal (the bottom panel). The same colors are assigned to
SiPMs close to the corresponding temperature sensors. To avoid overlap the red graph is shifted up by 5 units 
and the green one by 10.\label{fig:temp}}
\end{figure}

Despite of efforts to stabilize the temperature of the detector it varies at
a level of $1^\circ$C. A SiPM breakdown voltage has a strong temperature dependence~\cite{Gundacker}.
As a consequence other properties like gain, crosstalk, afterpulses probability and photon detection
efficiency (PDE) also have strong temperature dependence.
The detector is equipped with 36 temperature sensors, placed uniformly on the two sides of the copper
frame close to SiPM. Readout of the temperatures is done by an independent slow monitor system
every 5 minutes and is written separately from the main DAQ data files. Graphs of the temperatures are presented in
Figure~\ref{fig:temp} (top). At the bottom part of this figure single pixel signal 
dependence of the nearby SiPM is given. A clear anticorrelation is seen.

\begin{figure}[ht]
\includegraphics[width=\textwidth]{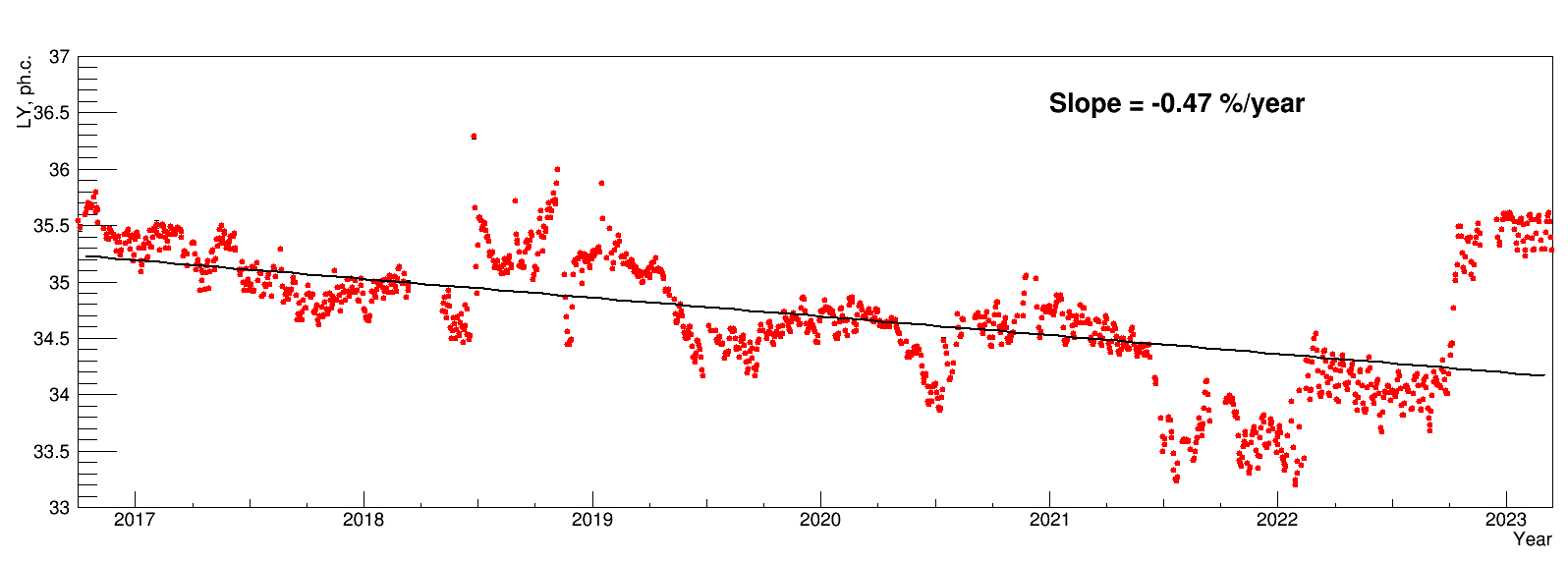}\\
\includegraphics[width=\textwidth]{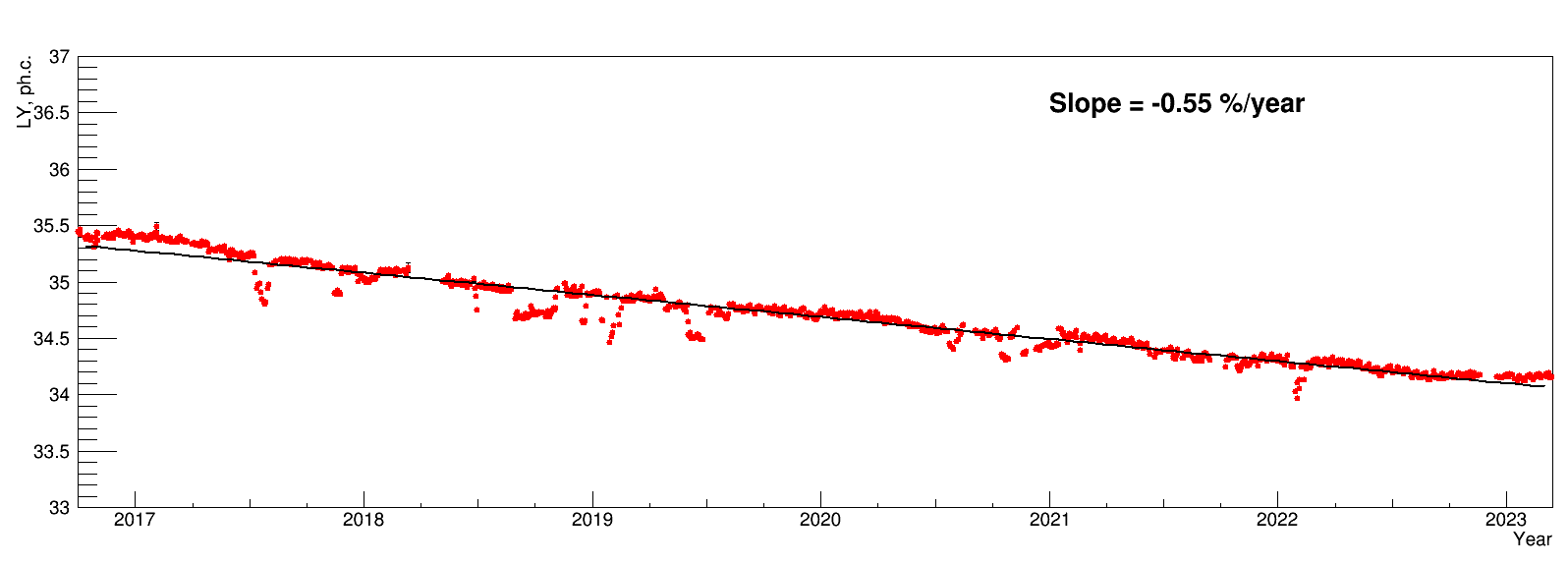}
\caption{Light yield dependence over calendar time without the temperature correction (top) and with the correction 
according to the equation (\ref{eq:comp})(bottom).
Statistical errors only.\label{fig:ageing}}
\end{figure}

The data used in this analysis is taken in the form of ph.c., so variation in the single pixel signal and crosstalk 
due to temperature is already taken into account, but changes in PDE are still not compensated. This compensation is
done in a simple linear approximation based on the single pixel signal measured for each SiPM in each data file. In
this approximation the corrected value of the signal is given by an equation:
\begin{equation}
    S^{ijk}_\mathrm{corr} = S^{ijk} (1 - \alpha \frac{A^{jk} - A^{k}_\mathrm{avg}}{A^{k}_\mathrm{avg}}) \:,
    \label{eq:comp}
\end{equation}
where $S$ is a number of ph.c., upper indexes correspond to event number ($i$), file number ($j$) and strip number ($k$). $A$ is a single
pixel signal and subscript "avg" denotes its average over all files. $\alpha$ is a correction coefficient common for all 
SiPM channels. The value of $\alpha = 0.85$ corresponds to the minimum of $\chi^2$ deviation from a linear
fit to LY over time dependence. The LY dependence over time is shown in 
Figure~\ref{fig:ageing}. The top panel
shows LY without the temperature correction. The corrected result is shown at the bottom. It 
is seen that some uncompensated effects are still present. Nevertheless the LY decrease 
with time is clearly seen. 

\begin{figure}[ht]
\centerline{\includegraphics[width=0.5\textwidth]{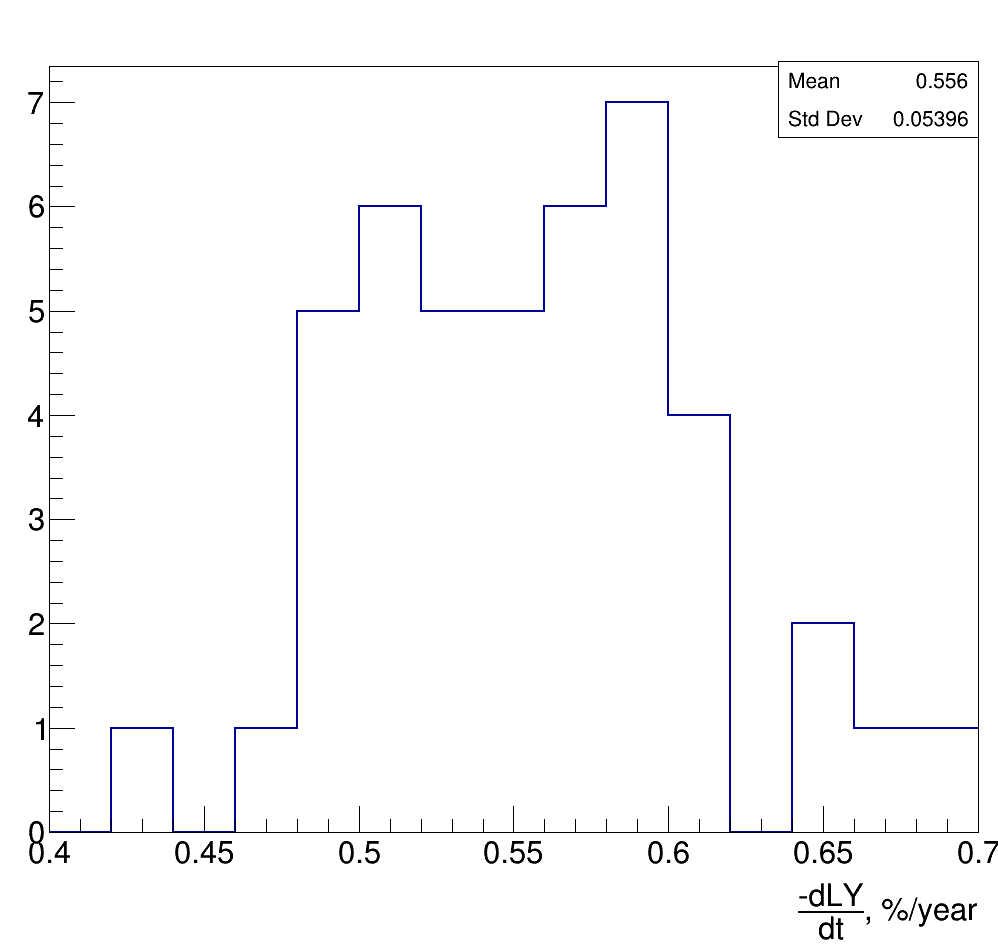}}
\caption{Light yield decay distribution for sections of the detector.\label{fig:50age}}
\end{figure}

Statistical only errors are shown in Figure~\ref{fig:ageing}. In order to estimate systematical errors
the detector was split into 50 sections of 50 neighbor strips each (5 horizontally strips to 10 rows of the same direction). Annual LY decrease was
calculated in these sections. The distribution of the results is shown in Figure~\ref{fig:50age}.
RMS of this distribution gives a good estimate of the systematic error, so the ageing of the whole
detector is $0.55 \pm 0.05(\mathrm{syst.})$~\%/year.

\begin{figure}[ht]
\centerline{\includegraphics[width=0.5\textwidth]{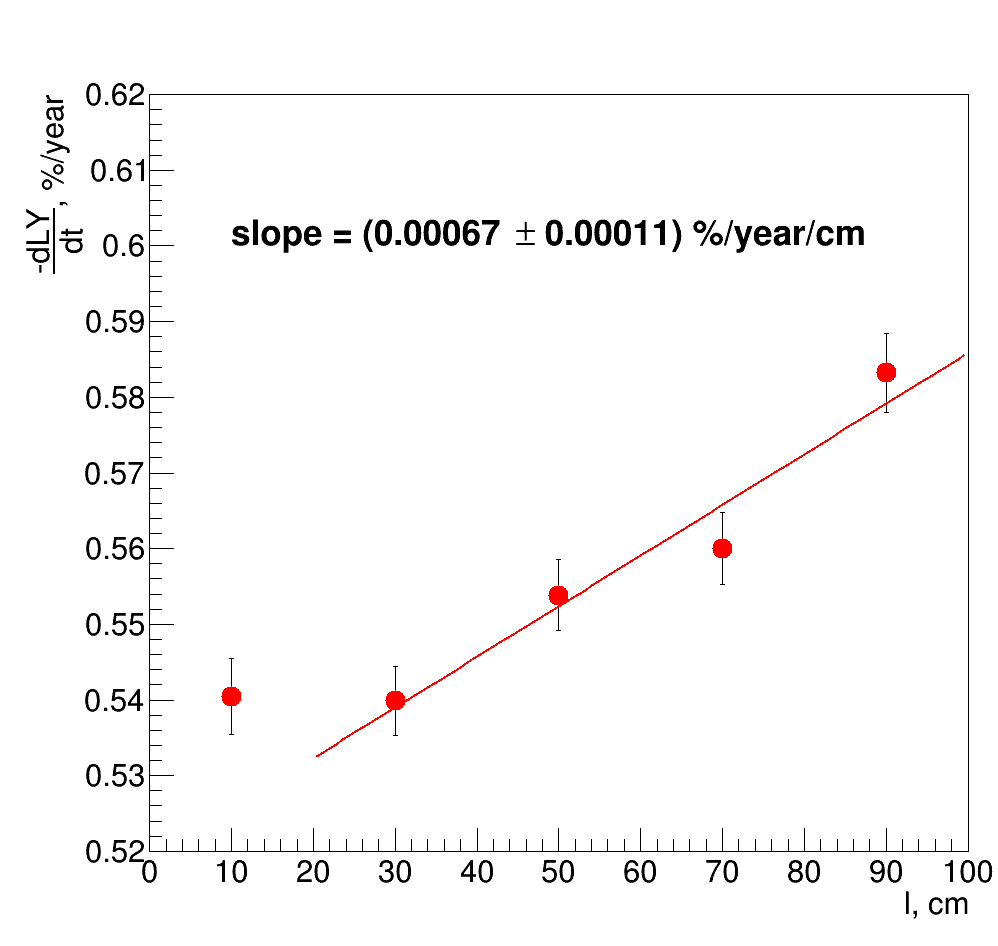}}
\caption{Light yield decrease dependence over the length along the strip. The line presents the linear fit
in the range $20 \div 100$~cm. \label{fig:WLS}}
\end{figure}

An observed decrease of the LY could have several reasons in addition to the degradation of the
scintillation material. They include ageing of WLS fibers, optical gel used to glue fibers and light receivers\footnote{No
ageing has been reported for SiPM unless in severe radiation conditions~\cite{Suna,Dolgoshein}}. 
We can not separate these factors with an exception of WLS fiber light attenuation length decrease, 
which can be seen as a larger ageing effect for longer distance along the strip. The LY decrease dependence over
the length along the strip is shown in Figure~\ref{fig:WLS}. A hint of ageing effect increase with the 
distance to a SiPM is seen. At moderate distances from SiPMs the light attenuation
is described by an exponent with the attenuation length $L_{att} = 3.9$~m. The 
observed increase in the ageing effect with the distance could be translated into the annual decrease of 
the attenuation length with a formula:
\begin{equation}
\frac{1}{L_{att}}\frac{\partial L_{att}}{\partial t} = L_{att} \frac{\partial}{\partial l}\left(\frac{1}{LY}\frac{\partial LY}{\partial t}\right) \: ,
\end{equation}
where $\frac{1}{LY}\frac{\partial LY}{\partial t}$ is the relative annual decrease of the 
LY and $l$ is the length from muon track to SiPM. 
The measured WLS attenuation length decrease, calculated from a linear fit in the range $20 \div 100$~cm
shown in Figure~\ref{fig:WLS}, is $0.26 \pm 0.04(\mathrm{stat.})$~\%/year.

\section{Conclusions}

A continuous measurement of the DANSS detector scintillator performance was carried out for 6.5 years. 
An annual LY decrease  $0.55 \pm 0.05(\mathrm{syst.})$~\%/year is clearly seen though it is smaller than
in other experiments. T2K measurement is in the range of $0.9 \div 2.2$~\%/year for different
detectors~\cite{T2K}, MINOS, which was using a very similar scintillating material, 
observed a decrease of about 2~\%/year~\cite{MINOS}. The small decrease of LY at DANSS
could be explained by the rather stable and mild environment of the DANSS detector. 
A contrary example is the fast $7 \div 10$~\%/year ageing of 
the MINERvA~\cite{MINERvA} scintillation detectors,
which are operated at a higher temperature ($27^{\circ}$C). Anyway ageing of DANSS scintillation 
strips is small and does not require special care in the data analysis.

\acknowledgments
DANSS collaboration is grateful to the directorates of ITEP and JINR for constant support of this work. 
The collaboration appreciates the permanent assistance of the KNPP administration and Radiation Safety 
Department staff.

The detector DANSS was created with the support of the State Atomic Energy Corporation Rosatom (ROSATOM) via state
contracts H.4x.44.90.13.1119 and H.4x.44.9B.16.1006. Long time operation of the experiment and initial data
analysis were supported by the Russian Science Foundation grant 17-12-01145 (2017--2021). The presented analysis 
as well as current detector operation are supported by the Russian Science Foundation 
grant 23-12-00085 (\url{https://rscf.ru/project/23-12-00085/}) (2023--2025).


\begin{thebibliography}{99}
\bibitem{DANSS_JINST} DANSS Collaboration. (I.~Alekseev {\em et al.}), {\it JINST} {\bf 11} (2016) P11011
\bibitem{DANSS_PLB} DANSS Collaboration. (I.~Alekseev {\em et al.}), {\it Phys. Lett. B} {\bf 787} (2018) 56
\bibitem{Barnaby1962} C.~F.~Barnaby and J.~C.~Barton, {\it J. Sci. Instruments} {\bf 39} (1962) 176
\bibitem{Kambour1973} R.~P.~Kambour, {\it J. Polym. Sci. Macromol. Rev.} {\bf 7} (1973) 1
\bibitem{Sword2017} E.~D.~Sword, In Proceedings of the IEEE International Symposium on Technologies for 
    Homeland Security (HST) 2017 (2017) 1
\bibitem{Yousif2013} E.~Yousif and R.~Haddad, Springerplus {\bf 2} (2013) 398
\bibitem{T2K} T2K Collaboration. (K.~Abe {\em et al.}), {\it JINST} {\bf 17} (2022) P10028
\bibitem{MINOS} MINOS Collaboration, (D.~G.~Michael {\em et al.}), 
    {\it Nucl. Instrum. Meth. Phys. Res. A} {\bf 596} (2008) 190
\bibitem{MINERvA} MINERvA Collaboration (L.~Aliaga {\em et al.}), 
    {\it Nucl. Instrum. Meth. Phys. Res. A} {\bf 743} (2014) 130
\bibitem{RepStrip} P.~N.~Zhmurin {\em et al.}, Development of Manufacturing Technology, Production and 
    Delivering of Plastic Strip Neutron Detectors. Report on Research and Development Work. Institute of 
    Scintillating Materials Ukraine National Academy of Science: Kharkiv, Ukraine, 2008. (In Russian)
\bibitem{KURARAY} Wavelength Shifting Fibers from Kuraray Co., Ltd., Tokyo, Japan
    \url{http://kuraraypsf.jp/psf/ws.html}
\bibitem{SUREL} SUREL-SL1, \url{http://www.surel.ru/silicone/76/} (In Russian)
\bibitem{WFD} I.~G.~Alekseev {\em et al.}, {\it Instruments Exp. Tech.} {\bf 61} (2018) 349
\bibitem{CalibPaper} I.~G.~Alekseev {\em et al.}, {\it Phys. Part. Nucl. Lett.} {\bf 15} (2018) 272
\bibitem{DANSS_muon} DANSS Collaboration. (I.~Alekseev {\em et al.}), {\it Eur.Phys.J.C} {\bf 82} (2022) 515
\bibitem{Gundacker} Stefan Gundacker and Arjan Heerin, {\it Phys. Med. Biol.} {\bf 65} (2020) 17TR01
\bibitem{Suna} Y. Sun and J. Maricic, {\it JINST} {\bf 11} (2015) C01078
\bibitem{Dolgoshein} B. Dolgoshein {\em et al.}, {\it Nucl. Instrum. and Meth. Phys. Res. A} {\bf 563} (2006) 368
\end{thebibliography}
\end{document}